\documentclass[%
 aip,
 amsmath,amssymb,
 reprint,%
]{revtex4-2}

\usepackage[dvips]{graphicx}
\usepackage{dcolumn}
\usepackage{bm}

\usepackage[utf8]{inputenc}
\usepackage[T1]{fontenc}
\usepackage{mathptmx}
\usepackage{etoolbox}
\usepackage{xcolor}

\renewcommand{\Re}{\mathop{\mathrm{Re}}}
\renewcommand{\Im}{\mathop{\mathrm{Im}}}
\newcommand{\Tr}{\mathop{\mathrm{Tr}}}

\makeatletter
\def\@email#1#2{%
 \endgroup
 \patchcmd{\titleblock@produce}
  {\frontmatter@RRAPformat}
  {\frontmatter@RRAPformat{\produce@RRAP{*#1\href{mailto:#2}{#2}}}\frontmatter@RRAPformat}
  {}{}
}%
\makeatother

\begin{document}

\preprint{AIP/123-QED}

\title{Single-junction quantum-circuit refrigerator}
\author{V. Vadimov}
\affiliation{QCD Labs, QTF Centre of Excellence, Department of Applied Physics, Aalto University, P.O. Box 13500, FI-00076 Aalto, Espoo, Finland}
\affiliation{MSP Group, QTF Centre of Excellence, Department of Applied Physics, Aalto University, P.O. Box 11000, FI-00076 Aalto, Espoo, Finland}
\affiliation{Institute for Physics of Microstructures, Russian
Academy of Sciences, 603950 Nizhny Novgorod, GSP-105, Russia}

\author{A. Viitanen}
\affiliation{QCD Labs, QTF Centre of Excellence, Department of Applied Physics, Aalto University, P.O. Box 13500, FI-00076 Aalto, Espoo, Finland}

\author{T. M\"orstedt} 
\affiliation{QCD Labs, QTF Centre of Excellence, Department of Applied Physics, Aalto University, P.O. Box 13500, FI-00076 Aalto, Espoo, Finland}

\author{T. Ala-Nissila}
\affiliation{MSP Group, QTF Centre of Excellence, Department of Applied Physics, Aalto University, P.O. Box 11000, FI-00076 Aalto, Espoo, Finland}

\affiliation{Interdisciplinary Centre for Mathematical Modelling, Department of Mathematical Sciences, Loughborough University, Loughborough LE11 3TU, UK}

\author{M. M\"ott\"onen}
\affiliation{QCD Labs, QTF Centre of Excellence, Department of Applied Physics, Aalto University, P.O. Box 13500, FI-00076 Aalto, Espoo, Finland}
\affiliation{VTT Technical Research Centre of Finland Ltd., QTF Center of Excellence, P.O. Box 1000, FI-02044 VTT, Finland}

\date{\today}

\begin{abstract}
We propose a quantum-circuit refrigerator (QCR) based on photon-assisted quasiparticle tunneling through a single normal-metal--insulator--superconductor (NIS) junction. In contrast to previous works with multiple junctions and an additional charge island for the QCR, we galvanically connect the NIS junction to an inductively shunted electrode of a superconducting microwave resonator making the device immune to low-frequency charge noise. At low characteristic impedance of the resonator and parameters relevant to a recent experiment, we observe that a semiclassical impedance model of the NIS junction reproduces the bias voltage dependence of the QCR-induced damping rate and frequency shift. For high characteristic impedances, we derive a Born--Markov master equation and use it to observe significant non-linearities in the QCR-induced dissipation and frequency shift. We further demonstrate that in this regime, the QCR can be used to initialize the linear resonator into a non-thermal state even in the absence of any microwave drive.
\end{abstract}

\maketitle


Superconducting quantum circuits~\cite{Nakamura1999, Wallraff2004, Krantz2019, Wallraff2021} have emerged as a highly promising platform for quantum simulations and quantum information processing. As these circuits advance in terms of performance and complexity, active control of individual devices within the circuit constitutes a key challenge. In particular, the control of temperature and dissipation in devices such as transmission lines~\cite{partanen_quantum-limited_2016}, microwave resonators~\cite{sevriuk_fast_2019} or qubits~\cite{hsu_tunable_2020, Yoshioka2021fast, basilewitsch_2020_fundamental, magnard_fast_2018, geerlings_demonstrating_2013} 
enables targeted cooling and initialization of these devices. This has already been achieved with the invention of the quantum-circuit refrigerator (QCR)~\cite{tan2017quantum}, a stand-alone, on-chip device that can locally cool superconducting circuits based on photon-assisted tunneling utilizing a pair of parallel normal-metal--insulator--superconductor (NIS) junctions~\cite{moerstedt2021recent, silveri_theory_2017}. We note that the NIS junctions have been widely used for refrigeration of electronic subsystem~\cite{giazotto2006opportunities,kuzmin2019photon,nguyen2013trapping,pekola2004cryogenics, muhonen2012micrometre} due to the non-linear Peltier effect predicted in Ref.~\onlinecite{bardas1995peltier}. Contrary to this, the QCR cools down the photonic degrees of freedom of the system.

In this work we present a simplified quantum-circuit refrigerator based on a single NIS junction and analyze this device in the case of direct coupling to a coplanar waveguide (CPW) resonator. The basic operation principle is similar to that of a double-junction refrigerator: A small bias voltage $V_{0}<\Delta /e$, where $\Delta$  is the superconductor gap parameter, is applied across the junction. Quasiparticles can tunnel through the insulating barrier, energetically allowed by absorption of photons from the coupled circuit which decreases the temperature of the electric degrees of freedom of the circuit~\cite{silveri_theory_2017, giazotto_opportunities_2006, courtois2014electronic, lowell2013macroscale}. Such tunneling events induce quantum-state transitions, for example, in a microwave resonator~\cite{ingold1992charge, sevriuk_fast_2019}. With increasing voltages $V_{0}>\Delta /e$, photon emission rate into the circuit approaches that of photon absorption, which leads to heating~\cite{masuda_observation_2018, hyyppa_calibration_2019}. Therefore, we operate the QCR in the range $\Delta-\hbar \omega_\textrm{r}<eV_0<\Delta$, with $\omega_\textrm{r}$ being the angular frequency of the lowest mode of the circuit.

We first present the design of a single-junction QCR coupled to a resonator, compare it to conventional double-junction microcooler devices, and analyze this system in terms of its tunable Lamb shift and dissipation rate. We further investigate the case of a single-junction QCR coupled to a high-impedance resonator, including multiphoton processes. Finally, we present our conclusions and explore further theoretical and experimental opportunities with this device.

\begin{figure}
    \centering
    \includegraphics[width=\columnwidth]{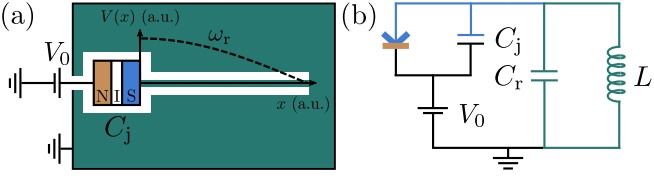}
    \caption{(a) Design of a single-junction QCR galvanically coupled to a $\lambda/4$ CPW resonator. The voltage profile of the fundamental mode at angular frequency $\omega_\mathrm{r}$ has an anti-node at the superconductor lead and a node where the center conductor connects to the ground plane (green). The NIS junction of capacitance $C_\mathrm{j}$ is biased with a voltage $V_0$ applied to the normal-metal lead. (b) Lumped-element circuit model of the fundamental resonator mode coupled to the NIS junction.}
    \label{fig:design}
\end{figure}

The system under study in Fig.~\ref{fig:design} consists of a single normal-metal--insulator--superconductor tunnel junction galvanically coupled to a superconducting resonator. The normal-metal lead is used to voltage bias the junction, providing on-demand control of tunneling. This novel implementation of a QCR differs from the previous devices~\cite{tan2017quantum} in two ways. First, we employ a single NIS junction as opposed to a double junction in a SINIS structure. Second, we avoid forming a charge island between the junction and the resonator by galvanically coupling the resonator to the NIS junction as opposed to capacitive coupling. The primary advantage of this approach is to mitigate low-frequency charge noise caused by changes in the charge state of the island and of the charge traps in its vicinity~\cite{hsu2021charge}. To achieve an equal tunneling rate and cooling power as for the double-junction device~\cite{tan2017quantum, silveri_theory_2017}, the corresponding single-junction QCR can be designed with half the tunneling resistance. In an experimental setting, the single-junction QCR can be controlled with a single line, whereas the double-junction QCR is typically operated through two separate control lines, an input and an output line~\cite{tan2017quantum, silveri2019broadband, sevriuk_fast_2019}. For the single-junction QCR, a galvanically coupled $\lambda/4$ CPW resonator naturally fixes the superconducting electrode to the ground potential at low frequencies.

We begin with an analysis of the dissipation and frequency shifts induced by the coupling of the resonator to the NIS junction. To this end, we employ classical equations of motion for the electromagnetic field in the circuit. The current-voltage characteristic of an NIS junction is both strongly non-linear and non-local in time. However, if we assume the ac component of the voltage to be small,
we can replace the junction by an effective frequency-dependent conductance, which can be controlled by the dc component of the voltage. To proceed, we define
\begin{gather}
    V_\textrm{NIS}(t) = V_\textrm{dc} + \frac{1}{2\pi} \int \limits_{-\infty}^{+\infty} V_\textrm{ac}(\omega) \textrm{e}^{-\textrm{i}\omega t}\;\mathrm d\omega, \\
    I_\textrm{NIS}(t) \approx I_\textrm{dc}(V_\textrm{dc}) + \frac{1}{2\pi} \int\limits_{-\infty}^{+\infty} G_\textrm{NIS}(V_\textrm{dc}, \omega) V_\textrm{ac}(\omega) \textrm{e}^{-\textrm{i}\omega t}\;\mathrm d\omega,
    \label{eq:NIS_current}
\end{gather}
where $G(V_\textrm{dc},\omega)$ is the conductance of the junction at bias voltage $V_\textrm{dc}$ and angular frequency $\omega$. Such an expansion is valid in the limit~\cite{tucker1985quantum} $|e V_\textrm{ac}(\omega)| \ll |\hbar \omega|$.

In order to find the dissipation rate and the Lamb shift of the resonator caused by the coupling to the NIS junction we write the Kirchhoff's law for the circuit shown in Fig.~\ref{fig:design}(b) as
\begin{equation}
    I_\textrm{NIS}[V_0 - \dot \varphi] = C \ddot \varphi + \frac{\varphi}{L} ,
\end{equation}
where $C = C_\textrm{j} + C_\textrm{r}$, 
\begin{equation}
    \varphi(t) = \int \limits_{-\infty}^{t} V(t')\;\mathrm dt',
\end{equation}
and $V(t)$ is the voltage across the capacitance $C_\textrm{r}$. Employing Eq.~\eqref{eq:NIS_current} and going to the Fourier picture we obtain an equation for the eigenfrequency of the resonator:
\begin{equation}
    -\omega^2 C - \mathrm{i} \omega G_\textrm{NIS}(V_0, \omega) + \frac{1}{L} = 0
\end{equation}
Provided the conductance is small $G_\textrm{NIS}(V_0, \omega) \ll Z_\textrm{r}^{-1}$ and changes smoothly on the frequency scale of $G_\mathrm{NIS}(V_0, \omega_{\mathrm r}) Z_\textrm{r} \omega_{\mathrm r}$, where $Z_\textrm{r} = \sqrt{L / C}$ is the characteristic impedance of the resonator, we can find the angular-frequency shift $\omega_\textrm{L}$ and the dissipation rate $\gamma$ as
\begin{equation}
    \omega_\textrm{L} - \mathrm{i} \gamma \approx -\frac{\mathrm{i}}{2}{G_\textrm{NIS}(V_0, \omega_\textrm{r})} \omega_\textrm{r} Z_\textrm{r},
    \label{eq:lamb_and_dissipation} 
\end{equation}
where $\omega_\textrm{r} = 1 / \sqrt{LC}$ is the resonance frequency of the bare $LC$ circuit.
Thus, both of these quantities are proportional to the imaginary and real parts of the NIS junction conductance. We can use this simple approach when the voltage across the capacitor satisfies the condition~$|e V(\omega)| \ll |\hbar \omega|$. In the quantum regime, voltage fluctuations are bounded from below by zero point fluctuations~\cite{blais2021circuit} $\delta V(\omega) / |\omega| \sim \sqrt{\hbar Z_\textrm{r} / 2}$. Thus, we obtain the applicability criterion for our result to be $\sqrt{{\pi Z_\textrm{r}}/{R_\textrm{K}}} \ll 1$, where $R_\textrm{K} = 2 \pi \hbar /  e^2$ is the von Klitzing constant.

The real and the imaginary parts of the junction conductance are given by the following expressions~\cite{tucker1985quantum}:
\begin{multline}
    \Re \left[ G_\textrm{NIS}(V_0, \omega)\right] = \frac{1}{8 R_\textrm{j}\omega} \int \limits_{-\infty}^{+\infty} \Im \left[\frac{\hbar \omega' + \textrm{i} \gamma_\textrm{D}\Delta}{\sqrt{\Delta^2 - (\hbar \omega' + \textrm{i} \gamma_\textrm{D} \Delta)^2}}\right]\\ \left\{\tanh \left[\frac{\hbar (\omega' + \omega) - e V_0}{2 k_\textrm{B} T}\right] + \tanh\left[\frac{\hbar (\omega' + \omega) + e V_0}{2 k_\textrm{B} T}\right] - \right . \\ \left. \tanh \left[\frac{\hbar (\omega' - \omega) - e V_0}{2 k_\textrm{B} T}\right] - \tanh\left[\frac{\hbar (\omega' - \omega) + e V_0}{2k_\textrm{B} T}\right]
    \right\}\;\mathrm d\omega',
    \label{eq:real_conductance}
\end{multline}
\begin{multline}
    \Im  \left[G_\textrm{NIS}(V_0, \omega) \right] = \frac{1}{4 R_\textrm{j}\omega} \int\limits_{-\infty}^{+\infty} \Re \left[\frac{\hbar \omega' + \textrm{i} \gamma_\textrm{D}\Delta}{\sqrt{\Delta^2 - (\hbar \omega' + \textrm{i} \gamma_\textrm{D} \Delta)^2}}\right]\\
    \left(
        \frac{1}{2} \left\{\tanh\left[\frac{\hbar (\omega' + \omega) - e V_0}{2k_\textrm{B} T}\right] +\tanh\left[\frac{\hbar (\omega' + \omega) + e V_0}{2k_\textrm{B} T}\right]+ \right. \right. \\ \left. \left . 
        \tanh\left[\frac{\hbar (\omega' - \omega) - e V_0}{2k_\textrm{B} T}\right] +\tanh\left[\frac{\hbar (\omega' - \omega) + e V_0}{2k_\textrm{B}T}\right]
        \right\} - \right . \\ \left.
        \tanh \left[\frac{\hbar \omega' - eV_0}{2k_\textrm{B} T}\right] - \tanh \left[\frac{\hbar \omega' + e V_0}{2 k_\textrm{B} T}\right]
    \right)\;\mathrm d\omega',
    \label{eq:imaginary_conductance}
\end{multline}
where $R_\textrm{j}$ is the junction resistance in the normal state, $\gamma_\textrm{D}$ is the Dynes parameter of the junction, and $T$ is the electron temperature of the normal metal.
\begin{figure}
    \centering
    \includegraphics[width=\linewidth]{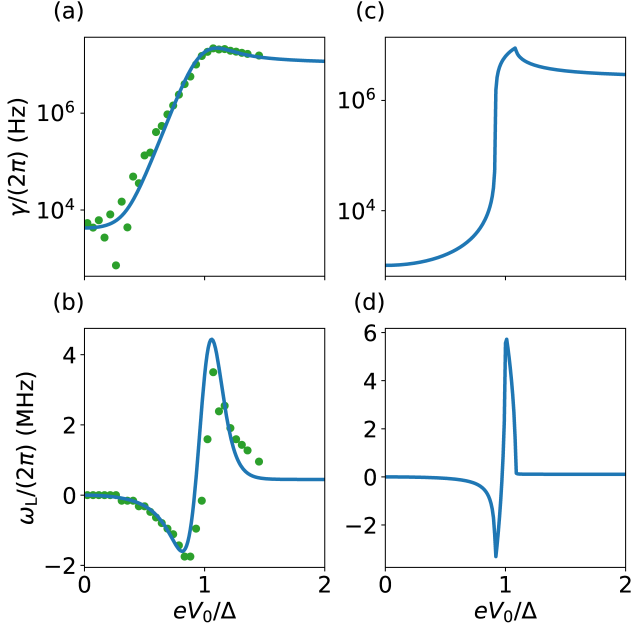}
    \caption{(a) QCR-induced dissipation rate and (b) frequency shift of a CPW resonator as functions of the bias voltage of the NIS junction. Blue lines correspond to Eqs.~\eqref{eq:lamb_and_dissipation}, \eqref{eq:real_conductance}, and \eqref{eq:imaginary_conductance}, the green dots show the experimental data from Ref.~\onlinecite{silveri2019broadband}. The QCR-induced frequency shift is defined to vanish at zero voltage $V = 0$. The device parameters are given by $\omega_\textrm{r} = 2\pi \times 4.67$~GHz, $Z_\textrm{r} =34.8$~$\mathrm \Omega$, $\Delta = 0.215$~meV, $T = 0.17$~K, $\gamma_\textrm{D} = 4\times 10^{-4}$, and the junction resistance $R_\textrm{j}=16$~$\mathrm{k \Omega}$ is obtained from a fit to the experimental relaxation rate. 
    Panels (c) and (d) show the dissipation rate and the frequency shift, respectively, for identical parameters except for $T = 0$~K. The positions of the dips and peaks in the frequency shifts are determined by $eV_0 = \Delta,~\Delta \pm \hbar \omega_\mathrm{r}$.}
    \label{fig:lamb_shift}
\end{figure}
The frequency shift and the dissipation rate of the resonator are shown in Fig.~\ref{fig:lamb_shift} for parameters the corresponding to the experiments in Ref.~\cite{silveri2019broadband} where a double-junction QCR capacitively coupled to a $\lambda/2$ resonator is employed. The theoretical results from Eqs.~\eqref{eq:lamb_and_dissipation}, \eqref{eq:real_conductance}, and \eqref{eq:imaginary_conductance} well agree with the experimental data despite the differences in the design of the sample and our model. Such a good correspondence can be explained by the fact that the frequency shift and the dissipation rate are both proportional to the sum of the conductances of the NIS junctions in the QCR. Provided that the SINIS junction is symmetric, the expressions for the frequency shift and the dissipation rate coincide with Eq.~\eqref{eq:lamb_and_dissipation} where the tunneling resistance and the characteristic impedance are adjusted accordingly. 

If the condition $\sqrt{\pi Z_\textrm{r} / R_\textrm{K}} \ll 1$ is violated, multi-photon processes become important in the quantum limit. In this case, we to quantize the electromagnetic field in the circuit. Employing the standard field quantization approach~\cite{ingold1992charge}, we obtain the following Hamiltonian for the circuit:
\begin{gather}
    \hat H = \hat H_\textrm{r} + \hat H_\textrm{NS} + \hat H_\textrm{t}; \\
    \hat H_\textrm{r} = \frac{\hat q^2}{2C} + \frac{\hat \varphi^2}{2L};
\end{gather}
\begin{multline}
    \hat H_\textrm{NS} = 
    \sum\limits_{k\sigma} \left(\xi_k^\textrm{n} - e V_0\right) \hat d_{k\sigma}^\dag \hat d_{k\sigma} + \sum\limits_{p\sigma} \xi_p^\textrm{s} \hat c_{p\sigma}^\dag \hat c_{p\sigma} + \\ \Delta \sum\limits_{p} \left(\hat c_{p\uparrow} \hat c_{\bar p \downarrow} + \textrm{h.c.}\right);
\end{multline}
\begin{equation}
\hat H_\textrm{t} =  \sum\limits_{kp\sigma} \left(\gamma_{kp} \hat d_{k\sigma}^\dag \hat c_{p\sigma} \textrm{e}^{\textrm{i} \frac{e \hat \varphi}{\hbar}} + \textrm{h.c.}\right) ,
\end{equation} 
where $\hat \varphi$ and $\hat q$ are the canonically conjugate $[\hat \varphi, \hat q] = \textrm{i} \hbar$ flux and charge operators of the $LC$ circuit, respectively, $\hat d_{k\sigma}$ is the annihilation operator of an electron in mode $k$ with spin projection $\sigma$ in the normal metal, $\xi_k^\textrm{n}$ is the energy of this mode with respect to the Fermi energy, $\hat c_{p\sigma}$ is the annihilation operator of an electron in the mode $p$ with spin projection $\sigma$ in the superconductor, $\xi_p^\textrm{s}$ is the energy of this mode with respect to the Fermi energy, and $\gamma_{kp}$ is the tunneling matrix element. 
We emphasize that the coupling between the electromagnetic and fermionic degrees of freedom occurs through the charge shift operator $\exp(\textrm{i}{e \hat \varphi}/{\hbar})$ in the tunneling Hamiltonian, since the tunneling of an electron through the junction is associated with a change of the capacitor charge by a single elementary charge. Introducing the notations
\begin{gather}
    \hat a = \frac{1}{\sqrt{2 \hbar Z_\textrm{r}}} \left(\hat \varphi + \textrm{i} Z_\textrm{r} \hat q\right);~\alpha = e \sqrt{\frac{Z_\textrm{r}}{2 \hbar}} = 
    \sqrt{\frac{\pi Z_\textrm{r}}{R_\textrm{K}}};\\
    \hat F = \textrm{e}^{\textrm{i} \alpha \left(\hat a + \hat a^\dag\right)};~\hat \Theta = \sum\limits_{kp\sigma} \gamma_{kp} \hat d_{k\sigma}^\dag \hat c_{p\sigma},
\end{gather}
we obtain a Redfield master equation~\cite{redfield1965theory} which governs the dynamics of the resonator density operator $\hat \rho$ as
\begin{multline}
    \frac{\mathrm d\hat \rho}{\mathrm d t} = \mathcal L (\hat \rho) =  -\frac{\textrm{i}}{\hbar} \left[\hat H_\textrm{r}, \hat \rho\right] -\\ \frac{1}{\hbar^2} \int \limits_0^{+\infty} \left\{
    \left[\hat F \hat F^\dag(-\tau) \hat \rho - \hat F^\dag(-\tau) \hat \rho \hat F\right] \left \langle\hat \Theta^\dag(\tau) \hat \Theta\right \rangle + \right. \\ \left.
    \left[\hat \rho \hat F^\dag(-\tau) \hat F - \hat F \hat \rho \hat F^\dag(-\tau)\right] \left \langle \hat \Theta \hat \Theta^\dag(\tau)\right \rangle + \right . \\ \left.
    \left[\hat F^\dag \hat F(-\tau) \hat \rho - \hat F(-\tau) \hat \rho \hat F^\dag\right] \left \langle \hat \Theta(\tau) \hat \Theta^\dag \right \rangle + \right. \\ \left.
    \left[\rho \hat F(-\tau)\hat F^\dag - \hat F^\dag \hat \rho \hat F(-\tau)\right] \left \langle \hat \Theta^\dag \hat \Theta(\tau)\right \rangle
    \right\}\;\mathrm d\tau,
    \label{eq:redfield}
\end{multline}
where $\mathcal L$ is the Liouvillian of the system, and the time dependent operators are given in the interaction picture:
\begin{gather}
    \hat F(\tau) = \textrm{e}^{\textrm{i} \frac{\hat H_\textrm{r} \tau}{\hbar}} \hat F  \textrm{e}^{-\textrm{i}\frac{\hat H_\textrm{r} \tau}{\hbar }},~
    \hat \Theta(\tau) = \textrm{e}^{\textrm{i} \frac{\hat H_\textrm{NS} \tau}{\hbar}} \hat \Theta\, \textrm{e}^{-\textrm{i}\frac{\hat H_\textrm{NS} \tau}{\hbar }}.
\end{gather}
Quantum-mechanical averaging is taken over the biased Fermi--Dirac distributions of the quasiparticles in the normal metal and in the superconductor~(see Supplementary Material (SM) for more information). 

Instead of solving Eq.~\eqref{eq:redfield} we simplify it further by a so-called secular approximation to the master equation, leaving only the terms which do not oscillate in time in the interaction picture. Thus the Liouvillian $\mathcal L$ can be replaced by its secular part $\mathcal L_\textrm{sec}$~(see SM for more information). The dynamics described by the original Liouvillian $\mathcal L$ can violate the non-negativity of the density operator and the secular approximation removes this inconvenience. Furthermore, the secular Liouvillian $\mathcal L_\textrm{sec}$ has a block structure, which significantly reduces the computational requirements. The validity of both the Born--Markov and secular approximations is determined by $\min(Z_\textrm{r}, R_\textrm{K}) / R_\textrm{j} \ll \min[1, k_\textrm{B} T / (\hbar \omega_\textrm{r})]$~(see SM for more information).

The steady-state density operator of the resonator satisfies $\mathcal L_\textrm{sec} (\hat \rho_0 )= 0$. Using the master equation we calculate the response function $D(\omega)$ with respect to an infinitesimal drive~\cite{albert2016geometry,ban2015linear,campos2016dynamical,villegas2016application} proportional to $\hat a + \hat a^\dag$:
\begin{equation}
    D(\omega) = -\textrm{i} \Tr \left\{\left(\hat a + \hat a^\dag\right) \left(\mathcal L_\textrm{sec} + \textrm{i} \omega\right)^{-1} \left[\hat a + \hat a^\dag, \hat \rho_0\right]\right\} .
\end{equation}
For an isolated resonator in its ground state, the response function equals
\begin{equation}
    D_0(\omega) = \frac{2 \omega_\textrm{r}}{\omega_\textrm{r}^2 - (\omega + \textrm{i}0)^2}.
\end{equation}
The poles of the response function as functions of the complex-valued frequency $\omega$ define the characteristic frequencies and decay rates of the modes in the system. From the definition of $D(\omega)$, we observe that these poles correspond to the eigenvalues $\lambda_{j}$ of the secular Liouvillian $\mathcal L_\textrm{sec}$. Accordingly, the general response function can be expanded as
\begin{equation}
    D(\omega) = \sum\limits_j \frac{f_j}{\lambda_j + \textrm{i} \omega} ,
\end{equation}
where the real and imaginary parts of $\lambda_{j}$ give the decay rate and frequency of the $j^{\rm th}$ transition, respectively. The complex amplitude $f_j$ quantifies how much this transition contributes to the response function.

\begin{figure}
    \centering
    \includegraphics[width=\linewidth]{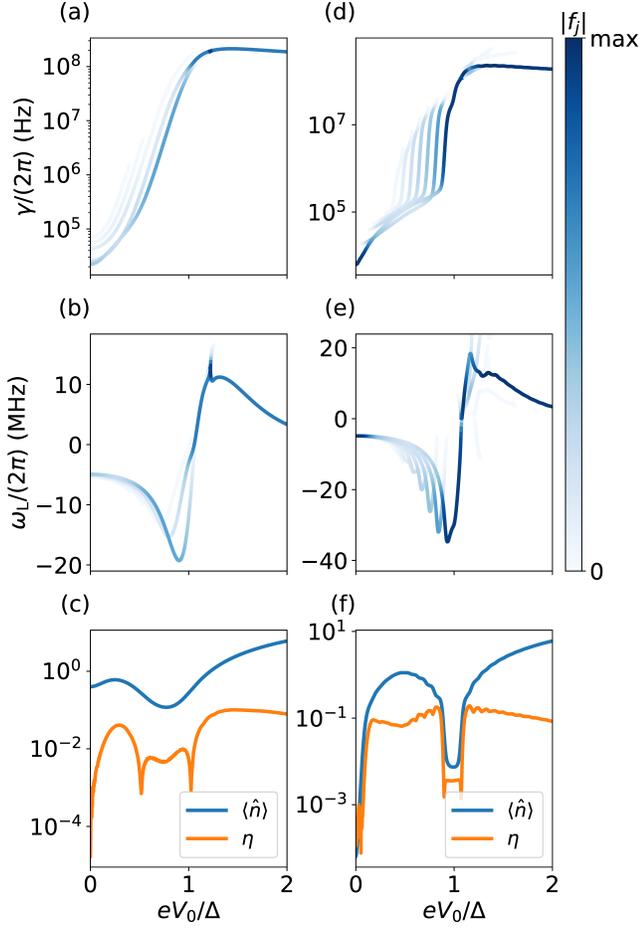}
    \caption{(a) QCR-induced dissipation rate, (b) frequency shift, and (c) the mean photon number $\langle \hat n \rangle$ and the distance to the closest thermal state $\eta$ for a high-impedance resonator as a function of the dc voltage across the NIS junction. The parameters are given by $\omega_\textrm{r} = 2\pi \times 4.67$~GHz, $Z_\textrm{r} =20$~k$\mathrm \Omega$, $\Delta = 0.215$~meV, $T = 0.17$~K, $\gamma_\textrm{D}= 4\times 10^{-4}$, $R_\textrm{j} = 640$~k$\mathrm \Omega$. (d)--(f) As panels (a)--(c) except for $T = 0.02$~K. The color of the lines in the panels (a), (b), (d), and (e) corresponds to the relative absolute value of $f_j$ which quantifies the contribution of the resonance to the response function $D(\omega)$.}
    \label{fig:multiphoton_lamb}
\end{figure}

Figure~\ref{fig:multiphoton_lamb} shows the dissipation rate and the frequency shift of the resonator as functions of the dc voltage applied to the NIS junction, obtained from eigenvalues of the Liouvillian of the master equation. 
Here, we study a high-impedance resonator with $Z_\textrm{r} = 20$~k$\mathrm \Omega$, and consequently increased the junction resistance to $R_\textrm{j} = 640$~k$\mathrm \Omega$ in order to work in the validity regime of the Born--Markov theory. In contrast to the low-impedance case, we observe here several resonances that differ in frequency and decay rate. Lowering the temperature to $T = 0.02$~K significantly increases the number of pronounced branches. We suggest that these branches correspond to the multi-photon transitions in the resonator, similar to those found in Ref.~\onlinecite{viitanen2021photon}. If the voltage is $eV_0 \approx \Delta - \hbar \omega_\textrm{r} n$, where $n$ is a positive integer, then tunneling processes with absorption of $n$ and more photons from the resonator are allowed. This implies that the resonator states with $n$ or more photons rapidly decay to lower-energy states, while states with less than $n$ photons decay relatively slowly. Thus, the system relaxes to some non-equilibrium steady state with almost zero probability of having $n$ or more photons, whereas the lower-energy states may have non-negligible occupation. If $n=1$, the QCR absorbs all photons from the resonator and the system relaxes toward the ground state. At negative $n$ (voltages above $\Delta / e$), photon emission processes come into play and we effectively heat the system. However, due to the complicated non-linear interaction between the resonator and the QCR, the photon distribution remains non-thermal.

In order to confirm this qualitative picture, we calculate two static properties of the steady state: the mean occupation number $\langle \hat n \rangle = \Tr \{\hat \rho_0 \hat a^\dag \hat a\}$ and trace distance to the closest thermal state~\cite{hillery1987nonclassical} $\eta = \inf_\beta \left\| \hat \rho_0 - \left(1 - e^{-\beta}\right) e^{-\beta \hat a^\dag \hat a}\right\|$, where $\|\hat A\| = \Tr \{\sqrt{\hat A \hat A^\dag}\}$. These quantities are shown in Figs.~\ref{fig:multiphoton_lamb}(c) and~\ref{fig:multiphoton_lamb}(f) as functions of the bias voltage $V_0$. At very low voltages the system is in a thermal state; hence $\eta$ is very low and $\langle \hat n\rangle$ corresponds to the thermal distribution with temperature $T$. With increasing voltage the resonator heats up due to the finite Dynes parameter of the superconductor and weak tunneling to the subgap states. When approaching the voltage $eV_0 \sim \Delta$, the occupation number decreases and so does the the distance to the thermal state, since single-photon processes come into play. This effect is especially pronounced in the low-temperature case shown in Fig.~\ref{fig:multiphoton_lamb}(f), where a sudden drop almost to the ground state in the interval $\Delta - \hbar \omega_\textrm{r} < eV_0 < \Delta + \hbar \omega_\textrm{r}$ is visible. With a further voltage increase, photon emission comes into play and the QCR starts to heat the resonator. Interestingly, the steady state in this regime still remains non-thermal for high-characteristic-impedance resonators.

To summarize, in this paper we have proposed a simple design of a QCR consisting of a single NIS junction galvanically coupled to the cooled system. We showed that the QCR-induced dissipation rate and frequency shift of the low-characteristic-impedance resonator can be found from a semiclassical treatment of the dynamics of the electromagnetic field in the circuit. In order to study the high-characteristic-impedance resonators and to obtain a distribution of photons in the steady state, we employed Born--Markov master equation approach to the resonator dynamics. We calculated the response function, the poles of which give the characteristic frequencies and decay rates of the resonator. We showed that a single bright resonance of the response function for the low-characteristic-impedance resonators splits into several resonances for the high-characteristic-impedance resonators, which correspond to the transitions between the shifted consecutive levels of the resonator. The short decay time of the high-energy levels due to the multi-photon transitions present in this system leads to a significant broadening of some of the transitions. We also calculated the mean photon number and the distance between the steady state to the closest thermal state. We showed that for the high-characteristic-impedance resonators the photon distribution function is essentially non-thermal due to the multi-photon transitions.
With this single-junction device we can achieve all the desired functionalities of a double-junction QCR. It can be used as a platform for studying open quantum systems which exhibit nontrivial phenomena in the non-linear 
limit.

\section*{Supplementary Material}
See Supplementary Material for the correlation functions of the NIS junction and the details of the secular approximation to the Redfield master equation.

\acknowledgments
We thank Alexander Mel'nikov, Sergei Sharov, Dmitry Golubev, Gianluigi Catelani, and Matti Silveri for useful discussions. This work was financially supported by the European Research Council under Grant No. 681311 (QUESS), Grant No. 957440 (SCAR), by the Academy of Finland under Grant No.~318937 and under its Centres of Excellence Program (projects 312300 and 312298), and by the Finnish Cultural Foundation. We acknowledge the computational resources provided by the Aalto Science-IT project.

\bibliography{bibliography,references_from_mikkos_zotero}

\end{document}


\preprint{AIP/123-QED}

\title{Supplementary material \\ Single-junction quantum-circuit refrigerator}
\author{V. Vadimov}
\affiliation{QCD Labs, QTF Centre of Excellence, Department of Applied Physics, Aalto University, P.O. Box 13500, FI-00076 Aalto, Espoo, Finland}
\affiliation{MSP Group, QTF Centre of Excellence, Department of Applied Physics, Aalto University, P.O. Box 11000, FI-00076 Aalto, Espoo, Finland}
\affiliation{Institute for Physics of Microstructures, Russian
Academy of Sciences, 603950 Nizhny Novgorod, GSP-105, Russia}

\author{A. Viitanen}
\affiliation{QCD Labs, QTF Centre of Excellence, Department of Applied Physics, Aalto University, P.O. Box 13500, FI-00076 Aalto, Espoo, Finland}

\author{T. Mörstedt} 
\affiliation{QCD Labs, QTF Centre of Excellence, Department of Applied Physics, Aalto University, P.O. Box 13500, FI-00076 Aalto, Espoo, Finland}

\author{T. Ala-Nissila}
\affiliation{MSP Group, QTF Centre of Excellence, Department of Applied Physics, Aalto University, P.O. Box 11000, FI-00076 Aalto, Espoo, Finland}

\affiliation{Interdisciplinary Centre for Mathematical Modelling, Department of Mathematical Sciences, Loughborough University, Loughborough LE11 3TU, UK}

\author{M. M\"ott\"onen}
\affiliation{QCD Labs, QTF Centre of Excellence, Department of Applied Physics, Aalto University, P.O. Box 13500, FI-00076 Aalto, Espoo, Finland}
\affiliation{VTT Technical Research Centre of Finland Ltd., QTF Center of Excellence, P.O. Box 1000, FI-02044 VTT, Finland}

\date{\today}

\maketitle

\section{NIS correlation functions}
The dynamical correlation functions of the electronic tunneling operator $\hat \Theta$ are given by the following expressions:
\begin{gather}
    \left \langle \hat \Theta^\dag(\tau) \hat \Theta\right\rangle = \frac{2 R_\mathrm{K}\hbar^2}{R_\mathrm{j}} g_\mathrm{n}^>(\tau) g_\mathrm{s}^<(-\tau) \mathrm e^{\mathrm i \frac{e V_0}{\hbar} \tau}; \\
    \left \langle \hat \Theta \hat \Theta^\dag(\tau) \right\rangle = \frac{2 R_\mathrm{K} \hbar^2}{R_\mathrm{j}} g_\mathrm{n}^<(\tau) g_\mathrm{s}^>(-\tau) \mathrm e^{\mathrm i \frac{e V_0}{\hbar} \tau},
\end{gather}
where
\begin{gather}
    g_{\mathrm n(\mathrm s)}^{>(<)}(\tau) = \frac{1}{2\pi} \int g_{\mathrm n(\mathrm s)}^{>(<)}(\omega) \mathrm e^{-\mathrm i\omega \tau}\;\mathrm d\omega;\\
    g_{\mathrm n}^{>(<)}(\omega) = \frac{\mathrm e^{-\frac{\omega^2}{\Omega^2}}}{1 + \mathrm e^{\mp\frac{\hbar \omega}{k_{\mathrm B} T}}};\\
    g_{\mathrm s}^{>(<)}(\omega) = \frac{\mathrm e^{-\frac{\omega^2}{\Omega^2}}}{1 + \mathrm e^{\mp\frac{\hbar \omega}{k_{\mathrm B} T}}}\Im 
    \frac{\hbar \omega + \mathrm i \gamma_{\mathrm D} \Delta}{\sqrt{\Delta^2 - (\hbar \omega + \mathrm i \gamma_{\mathrm D} \Delta)^2}},
\end{gather}
are the quasiclassical greater and lesser Green's functions of the normal metal and the superconductor, and $\Omega$ is an ultraviolet cut-off frequency introduced in order to make all the integrals convergent. It represents a finite band width of the superconductor and the normal metal and is the largest frequency scale in the system. We choose a Gaussian cut-off since its shape does not affect the low energy physics of the system. The value of the cut-off frequency, which we choose to be $\hbar \Omega = 10$~meV, does not affect the dynamics of the resonator in the low characteristic impedance regime, but may give logarithmic corrections to the dissipation rates and the Lamb shift for the high characteristic impedance resonators. The characteristic decay rate of the correlation functions is determined by the temperature and is equal to $k_{\mathrm B} T / \hbar$. This rate should be sufficiently higher than the relaxation rate of the resonator $\min(Z_{\mathrm r}, R_{\mathrm K}) / R_{\mathrm j} \omega_{\mathrm r}$ which gives us a restriction on temperature at which Born--Markov approximation is valid.\newpage

\section{Secular approximation}

The Redfield master equation can be written in the following way:
\begin{multline}
    \frac{\mathrm d\hat \rho}{\mathrm dt} = -\mathrm i\omega_{\mathrm r} \left[\hat a^\dag \hat a, \hat \rho\right] - \\ \left(
    \hat F \hat F^\dag_1 \hat \rho - \hat F_1^\dag \hat \rho \hat F + \hat \rho \hat F_2^\dag \hat F - \hat F \hat \rho \hat F_2^\dag + \right . \\ \left. 
    \hat F^\dag \hat F_2 \hat \rho - \hat F_2 \hat \rho \hat F^\dag + \hat \rho \hat F_1 \hat F^\dag - \hat F^\dag \hat \rho \hat F_1
    \right),
\end{multline}
where
\begin{gather}
    \hat F_1 = \frac{1}{\hbar^2} \int\limits_0^{+\infty} \hat F(-\tau) \left \langle \hat \Theta^\dag \hat \Theta(\tau)\right \rangle\;\mathrm d\tau;\\
    \hat F_2 = \frac{1}{\hbar^2} \int\limits_0^{+\infty} \hat F(-\tau) \left \langle \hat \Theta(\tau) \hat \Theta^\dag \right \rangle\;\mathrm d\tau.
\end{gather}
We go to the frame rotating with the bare oscillator frequency $\omega_{\mathrm r}$:
\begin{equation}
    \hat \rho(t) = \mathrm e^{-\mathrm i\omega_{\mathrm r} \hat a^\dag \hat a t} \tilde \rho(t) \mathrm e^{\mathrm i \omega_{\mathrm r}  \hat a^\dag \hat a t} .
\end{equation}
In this frame the density operator $\tilde \rho$ satisfies the following equation:
\begin{multline}
    \frac{\mathrm d\tilde \rho}{\mathrm dt} =
    -\hat F(t) \hat F^\dag_1(t) \tilde \rho + \hat F_1^\dag(t) \tilde \rho \hat F(t) - \\ \tilde \rho \hat F_2^\dag(t) \hat F(t) + \hat F(t) \tilde \rho \hat F_2^\dag(t) - \\ \hat F^\dag(t) \hat F_2(t) \tilde \rho + \hat F_2(t) \tilde \rho \hat F^\dag(t) - \\ \tilde \rho \hat F_1(t) \hat F^\dag(t) + \hat F^\dag(t) \tilde \rho \hat F_1(t),
\end{multline}
where
\begin{gather}
    \hat F(t) = \mathrm e^{\mathrm i\omega_{\mathrm r} \hat a^\dag \hat a t} \hat F \mathrm e^{-\mathrm i\omega_{\mathrm r} \hat a^\dag \hat a t};\\
    \hat F_1(t) = \mathrm e^{\mathrm i\omega_{\mathrm r} \hat a^\dag \hat a t} \hat F_1 \mathrm e^{-\mathrm i\omega_{\mathrm r} \hat a^\dag \hat a t};\\
    \hat F_2(t) = \mathrm e^{\mathrm i\omega_{\mathrm r} \hat a^\dag \hat a t} \hat F_2 \mathrm e^{-\mathrm i\omega_{\mathrm r} \hat a^\dag \hat a t}.
\end{gather}
We write down the equation for the matrix elements of $\tilde \rho_{mn} = \langle m | \tilde \rho|n\rangle$ in the basis of the eigenfunctions of harmonic oscillator:
\begin{widetext}
\begin{multline}
    \frac{\mathrm d\tilde \rho_{mn}}{\mathrm d t} = \sum\limits_{pq}\left[F^\dag_{1mp} \tilde \rho_{pq} F_{qn} \mathrm e^{\mathrm i\omega_r(m-p+q-n) t}-F_{mp} F_{1pq}^\dag \tilde \rho_{qn} \mathrm e^{\mathrm i\omega_r(m-q) t} + F_{mp}\tilde \rho_{pq} F_{2qn}^\dag \mathrm e^{\mathrm i\omega_r(m - p + q - n) t} - \tilde \rho_{mp} F_{2pq}^\dag F_{qn} \mathrm e^{\mathrm i\omega_r(p - n) t} + \right. \\ \left . F_{2mp} \tilde \rho_{pq} F^\dag_{qn} \mathrm e^{\mathrm i\omega_r(m-p+q-n)t} - F^\dag_{mp} F_{2pq} \tilde \rho_{qn} \mathrm e^{\mathrm i\omega_r(m-q)t} + F^\dag_{mp} \tilde \rho_{pq} F_{1qn} \mathrm e^{\mathrm i\omega_r(m-p+q-n) t} - \tilde\rho_{mp} F_{1pq} F^\dag_{qn} \mathrm e^{\mathrm i\omega_r(p-n)t} \right],
\end{multline}
\end{widetext}
where
\begin{gather}
    F_{mn} = \langle m | \hat F | n\rangle,~F_{mn}^\dag = \langle m | \hat F^\dag | n \rangle;\\
    F_{1mn} = \langle m | \hat F_1 | n\rangle,~F_{1mn}^\dag = \langle m | \hat F_1^\dag | n \rangle;\\
    F_{2mn} = \langle m | \hat F_2 | n\rangle,~F_{2mn}^\dag = \langle m | \hat F_2^\dag | n \rangle.
\end{gather}
The point of the secular approximation is to drop the terms which oscillate in the rotating frame, leaving only stationary terms. This is justified if the characteristic decay rate $\min(Z_{\mathrm r}, R_{\mathrm K}) / R_{\mathrm j} \omega_{\mathrm r}$ is well below the bare resonator frequency $\omega_{\mathrm r}$. Finally we obtain the following equation for the resonator density operator:
\begin{multline}
    \frac{\mathrm d\tilde \rho_{mn}}{\mathrm dt} =\\ \sum\limits_p \left\{ \tilde \rho_{p (p+n-m)}\left[F_{1mp}^\dag F_{(p+n-m)n} + F_{mp} F^\dag_{2(p+n-m)n} + \right .\right. \\ \left. \left. F_{2mp} F^\dag_{(p+n-m) n} + F^\dag_{mp} F_{1(p+n-m)n}\right] - \right. \\ \left .\tilde \rho_{mn} \left[F_{mp} F^\dag_{1pm} + F^\dag_{2np}F_{pn} + F^\dag_{mp}F_{2pm} + F_{1np}F^\dag_{pn} \right]\right\} .
\end{multline}
One can notice that the system of equations can be separated into a set of independent systems which bind only matrix elements $\tilde \rho_{mn}$ with fixed $m-n$. This significantly reduces the computational resources required for the calculations.